\documentclass[a4paper]{ifacconf}

\makeatletter
\def\verbatim@font{\normalfont\ttfamily\footnotesize}
\makeatother

\usepackage[utf8]{inputenc}
\usepackage{graphicx,amsmath} 
\usepackage[round]{natbib} 
\usepackage{booktabs}

\usepackage{icomma} 
\usepackage{amsmath,amssymb,amsfonts} 
\usepackage{verbatim}
\usepackage{graphicx} 
\usepackage{psfrag} 
\usepackage{float} 
\usepackage{algorithm}
\usepackage{algpseudocode}
\usepackage{multirow}
\usepackage{multicol, blindtext, graphicx}
\usepackage{color}
\usepackage{xcolor}
\usepackage{verbatim}
\def\portugues{1} 
\ifx\portugues\undefined
\def\portugues{0}
\fi

\if\portugues0
   \usepackage[english]{babel}
  \else
   \usepackage[spanish,brazil,english]{babel}
\fi

\usepackage[T1]{fontenc}

\usepackage[utf8]{inputenc}

\usepackage{ae}

\if\portugues1

\newdimen\slantmathcorr
\def\oversl#1{
    \setbox0=\hbox{$#1$}
    \slantmathcorr=\wd0
    \hskip 0.2\slantmathcorr \overline{\hbox to 0.8\wd0{%
            \vphantom{\hbox{$#1$}}}}
    \hskip-\wd0\hbox{$#1$}
}

\def\undersl#1{
    \setbox0=\hbox{$#1$}
    \slantmathcorr=\wd0
    \underline{\hbox to 0.8\wd0{%
            \vphantom{\hbox{$#1$}}}}
    \hskip-0.8\wd0\hbox{$#1$}
}

\begin{document}
\selectlanguage{brazil}
\begin{frontmatter}
\title{Simplificação da representação digital do mapa tenda por meio de ponto fixo polarizado}

\author[First]{Mariana Teixeira}
\author[First]{Nayara Pinto Basílio}
\author[First]{Davidson Lafitte Firmo}
\author[First]{Erivelton G. Nepomuceno}
\author[Second]{Janier Arias-Garcia}
\address[First]{GCOM - Grupo de Controle e Modelagem, Dept. de Eng. Elétrica, UFSJ - Universidade Federal de São João del-Rei, MG, \\(email: marianateixeira6497@gmail.com, nayabasilio@gmail.com, davidson@ufsj.edu.br, nepomuceno@ufsj.edu.br)}
\address[Second]{Departamento de Engenharia Eletrônica, UFMG - Universidade Federal de Minas Gerais, Belo Horizonte (email: janier-arias@ufmg.br)}

\maketitle


\selectlanguage{english}
\renewcommand{\abstractname}{{\bf Abstract:~}}
\begin{abstract}
Chaotic systems have been investigated in several areas of engineering. In control theory, such systems have instigated the emergence of new techniques as well, have been used as a source of noise generation. The application of chaotic systems as pseudo-random numbers has also been widely employed in cryptography. One of the central aspects of these applications in high performance situations, such as those involving a large amount of data (Big Data), is the response of these systems in a short period of time. Despite the great advances in the design of chaotic systems in analog circuits, it is perceived less attention in the optimized design of these systems in the digital domain. In this work, the polarized fixed point representation is applied to reduce the number of digital elements. Using this approach, it was possible to significantly reduce the number of logic gates in the subtraction operation. When compared to other works in the literature, it has been viable to reduce by 50 \% the number of elements per bit of the digital representation of the tent map. The chaoticity was evidenced with the calculation of the Lyapunov exponent. Histogram, entropy and autocorrelation tests were used satisfactorily to evaluate the randomness of the represented system.

\vskip 1mm
\selectlanguage{brazil}
{\noindent \bf Resumo}:  Sistemas caóticos têm sido investigados em diversas áreas da engenharia. Na teoria de controle, tais sistemas têm instigado o surgimento de novas técnicas, bem como, têm sido utilizados como fonte de geração de ruído. A aplicação de sistemas caóticos como números pseudo-aleatórios também tem sido largamente empregada em criptografia. Um dos aspectos centrais nessas aplicações em situações que exigem alto desempenho, como as que envolvem um grande volume de dados (Big Data),  é a resposta desses sistemas em um curto intervalo de tempo. Apesar de grandes avanços na concepção de sistemas caóticos em circuitos analógicos, percebe-se uma menor atenção no projeto otimizado desses sistemas no domínio digital. Neste trabalho, aplica-se a representação de ponto fixo polarizado para reduzir o número de componentes lógicos. Com tal emprego, foi possível reduzir significativamente o número de portas lógicas na operação de subtração. Quando comparado com outros trabalhos na literatura, foi possível reduzir em 50\% o número de componentes por bit da representação digital do mapa tenda. A caoticidade foi evidenciada com o cálculo do expoente de Lyapunov. Histograma, entropia e teste de autocorrelação foram empregados satisfatoriamente para avaliar a aleatoriedade do sistema representado.
\end{abstract}

\selectlanguage{english}


\selectlanguage{brazil}
\end{frontmatter}

\section{Introdução}
\selectlanguage{brazil}

Grande parte das técnicas avançadas de controle industriais são baseadas na identificação e modelagem de processos. Um ponto chave para identificação é a seleção de um sinal de excitação bem projetado, que permite obter dados úteis para o desenvolvimento de modelos precisos \citep{LiYao2006}. Um dos sinais de perturbação mais utilizados são as sequências binárias pseudo-aleatórias (PRBS). Estas são interessantes pois possuem um espectro de frequência relativamente largo e sua covariância se aproxima ao ruído branco \citep{Tan2013}.

Na literatura existem diversas metodologias para gerar um sinal PRBS, como o uso de portas \textit{XOR} e registradores de deslocamento \citep{Khani2013, Pesoshin2016}. Para garantir que o processo dinâmico é constantemente excitado pelo sinal de entrada, estas metodologias devem assegurar que o período do sinal PRBS é maior que o tempo de acomodação do sistema a ser identificado \citep{Aguirre2015}. Por isso, nas últimas décadas, o uso de PRBS baseados em  mapas discretos caóticos ganhou grande interesse. Estes sistemas, mesmo que a precisão finita acarrete eventualmente uma dinâmica periódica, conseguem preservar a dinâmica caótica dentro de um longo período \citep{Sreenath2018}.

Apesar de grandes avanços nos sistemas caóticos em circuitos analógicos \citep{Spr2011,Spr2000a}, percebe-se uma menor atenção no projeto otimizado desses sistemas no domínio digital \citep{NEPOMUCENO201962}. Muitas vezes a digitalização dos mapas caóticos e as técnicas de degradação do caos podem consumir um grande volume de componentes, diminuindo sua eficiência energética e desempenho.  Um aspecto relevante para PRBS aplicados em controle de processos industriais, é a implementação de algoritmos em hardware de forma a otimizar a relação de custo, desempenho e energia. Com isto, este trabalho apresenta a digitalização do mapa tenda usando uma representação numérica alternativa, baseada na polarização do formato padrão de ponto fixo \citep{IEE2008}. Ao polarizar, ou seja, tornar a representação afim por parte, foi possível reduzir significativamente o número de portas lógicas na operação de subtração.

A contribuição deste trabalho é evidenciada pela representação em ponto fixo polarizado com sua consequente redução do número de elementos lógicos por bit. Na comparação com outros trabalhos da literatura \citep{Sreenath2018,Khani2013}, foi possível reduzir em 50\% o número de elementos digitais por bit.  Com essa redução, abre-se espaço para implementação em linguagem de descrição de hardware (VHDL, por exemplo) como já tem sido feito em outros trabalhos \citep{Silva2017,MB2015,Ismail2017}. 

É apresentado também uma metodologia de perturbação de sistema muito simples, baseada na alteração do bit menos significativo do sistema, capaz de diminuir a degradação do caos. A caoticidade foi evidenciada com o cálculo do expoente de Lyapunov \citep{MN2016,RCD1993}. Histograma, entropia e teste de autocorrelação foram empregados satisfatoriamente para avaliar a aleatoriedade do sistema representado.

O restante deste trabalho está organizado da seguinte forma. Na Seção \ref{sec:fund} os conceitos fundamentais são apresentados. Em seguida, na Seção \ref{sec:met}, a metodologia deste trabalho é descrita. Os resultados são apresentados na Seção \ref{sec:res}, enquanto as considerações finais estão na Seção \ref{sec:con}.

\section{Fundamentação Teórica}
\label{sec:fund}

\subsection{Mapa tenda}
\label{mtenda}
O mapa tenda é descrito por uma função recursiva que, para uma condição inicial $x_0$, fornece uma saída de acordo com a Equação \eqref{eq:1}. Para manter a propriedade caótica, o parâmetro de controle é $\mu\approx 2$.
\begin{equation}
    x_{n+1} = \left  \{ \begin{array}{cc}
        \mu x_n,     & x_n<\frac{1}{2}   \\
        \mu (1-x_n), & \frac{1}{2}\leq x_n
        \\ 
\end{array} \right .
\label{eq:1}
\end{equation}

 O mapa tenda é relativamente simples e ainda sim apresenta um comportamento caótico sensível às condições iniciais. Como apresentado em \cite{Crampin1994}, a análise do mapa tenda é muito simplificada para representação binária, tornando seu uso vantajoso para aplicações em circuitos digitais.
 
 \subsection{Expoente de Lyapunov}

    O expoente de Lyapunov é um número utilizado para indicar a dinâmica caótica de um sistema, analisando sua sensibilidade às condições iniciais. Para sistemas discretos, o expoente de Lyapunov é dado pela Equação \eqref{eq:2} \citep{MN2016,Nep2014,NMAR2017}, 
 \begin{equation}
     \Lambda = \lim_{m\rightarrow \infty } \dfrac{1}{N}\sum_{n=1}^{m-1}\log_2 \left | \dfrac{dF(x_n)}{dx_n} \right |
     \label{eq:2}
 \end{equation} 
 em que $m$ é o número de iterações, $F(x_n)=x_{n+1}$, $N$ é a quantidade de elementos calculados para a série $F(x_n)$ e $n = 1,2,3...$ é a dimensão do sistema. Expoentes de Lyapunov positivos indicam divergência na trajetória e comportamento caótico.

 \begin{figure*}[!ht]
    \centering
    \includegraphics[width =0.8\linewidth]{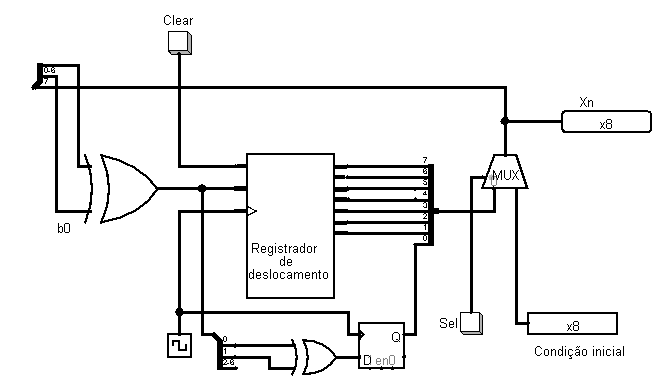}
    \caption{Circuito com 8 bits implementado no \textit{Logisim}. O circuito XOR a esquerda é composto por 7 portas \textit{XOR} que operam entre $b_0$ e $b_1$, $b_0$ e $b_2$, ..., $b_0$ e $b_7$ e são responsáveis por verificar se $x_n \geq \frac{1}{2}$ e, se necessário, fazer o complemento; O registrador de deslocamento realiza uma operação equivalente a multiplicação por 2; A porta \textit{XOR} na parte inferior é responsável pela perturbação e o multiplexador aplica a condição inicial ao sistema.}
    \label{fig:log}
\end{figure*}

 \subsection{Gerador de números pseudo aleatórios (\textit{PRNGs})}
 
 Para garantir o comportamento aparentemente aleatório de um \textit{PRNG}, a sequência gerada deve ser uniformemente distribuída e apresentar um período extremamente longo\citep{Wang2014}.
 
 Uma forma de avaliar a característica aleatória é a medida de entropia, que pode ser descrita como a medida de quão imprevisível é uma sequência e é definida por:
 \begin{equation}
     H(x)=-\sum_{i=1}^{n} p_i\log{p_i},
 \end{equation}
 onde $n$ é o tamanho da sequência, $x$ é a sequência e $p_i$ sua distribuição de probabilidade \citep{DelaFraga2017}.
 
 Quando $H(x)=1$, a distribuição da sequência é uniforme, sendo impossível prevê-la a partir dos dados anteriores. Se $H(x) = 0$ a sequência é totalmente previsível.
 
 O último teste a ser empregado para avaliar a pseudo-aleatoriedade do sistema é autocorrelação, que  é uma medida estatística que mostra o grau de associação entre valores anteriores e posteriores de um conjunto de dados \citep{Khani2013,Sreenath2018}, sendo uma ferramenta matemática útil para identificar a presença de períodos em um sistema. Um \textit{PRNG} deve apresentar autocorrelação muito próxima a zero para indicar a falta de padrões repetitivos.
 
 
 \section{Metodologia}
 \label{sec:met}
Como a Eq. \eqref{eq:1} somente apresenta o resultado desejado para $x_n \in [0,1]$, buscou-se simplificar a forma que um circuito lógico compreende números fracionários. Para tanto, foi adotado o padrão de ponto fixo fracionário descrito na Eq. \eqref{eq:3}, 
\begin{equation}\label{eq:3}
x_{10} \approx x_2 = 0,b_0 b_1 b_2 \ldots b_k,
\end{equation}
em que $ x_{10} $ é o número real a ser representado da base dez; $ x_2 $ é o número binário aproximado no sistema digital;  $ b $ representa o bit, ou valor lógico 0 e 1 e $ k $ representa o número de bits. Cumpre destacar que, como qualquer representação digital, o símbolo aproximado $ \approx $ pode dar lugar a igualdade para casos de ponto fixo, ou seja, números em que a conversão entre as bases é exata. Neste trabalho, foram investigadas representações com $ k \geq 8 $. Sabe-se que o $ b_k $ equivale a resolução, ou valor na última casa \textit{ulp}, com valor indicado como $ 2^{-k} $. A representação numérica de ponto fixo polarizado na base 2, indicada por $ {2^*} $ pode ser definida da seguinte forma:
\begin{equation}\label{key}
x_{2^*} = x_2 - ulp = x_2 - 2^{-k}.
\end{equation}
Como consequência, a representação do número $ 1_{10} $ em seu formato ponto fixo polarizado é descrita como
\begin{equation}
    1_{10}= 0,111...11_{2^*},
\end{equation}
e a subtração $(1-x_n)$ equivale a
\begin{equation}\label{sub}
1_{2^*}-x_n=\overline{x_n}
\end{equation}

Esta representação permite a simplificação da operação aritmética de subtração, como mostrado na Eq. \eqref{sub}. A fim de satisfazer a condição $x_n \geq \frac{1}{2}$ na Eq. \eqref{eq:1}, utiliza-se uma porta \textit{XOR} entre  $x_n = 0,b_0b_1b_2...b_k$ e o bit $b_0$, permitindo verificar a condição e fazer o complemento do número se ela for verdadeira. Ainda buscando reduzir o circuito digital, na Eq. \eqref{eq:1} utilizou-se $\mu = 2$ que, sendo potência de 2, permite realizar a multiplicação aritmética mediante apenas um registrador de deslocamento. Assim, digitalmente, o mapa tenda é representado como:
\begin{equation}
     x_{n+1} = \left  \{ \begin{array}{cc}
         2x_n  {\rm \ se}     & b_0=0,   \\
        2\overline{x_n} {\rm \  se} & b_0 = 1.
        \\ 
\end{array} \right .
\end{equation}
Como visto na seção \ref{mtenda}, o mapa tenda apresenta comportamento caótico para $\mu\approx2$. No entanto, quando $\mu = 2$, há uma degradação do caos e os resultados se tornam periódicos. Neste trabalho, para garantir as propriedades caóticas do sistema e diminuir este fenômeno de degradação causado principalmente pela representação numérica finita \citep{LCM2005,CLQL2015}, utilizou-se uma porta \textit{XOR} entre os últimos bits, de tal forma que
\begin{equation}
    b_k = b_k \oplus b_{k-1}
\end{equation}
é suficiente para causar uma perturbação. O circuito projetado foi montado no software \textit{Logisim}, em que é possível salvar em um arquivo \textit{.txt} os valores que foram registrados na notação binária. Com o auxílio do \textit{Matlab}, as representações binárias foram decodificadas para os valores decimais aproximados. Para verificar se o sistema digital caótico representou adequadamente o mapa tenda, foram feitas duas análises: o mapa de primeiro retorno e o cálculo do maior expoente de Lyapunov, conforme descrito em \citep{MN2016}. Para verificar a viabilidade do circuito como gerador pseudo aleatório, foram realizados os testes de entropia e autocorrelação descritos na Seção \ref{sec:fund}, além da avaliação do histograma.

\section{Resultados}
\label{sec:res}

O circuito projetado foi implementado no software \textit{Logisim}. A implementação para 8 bits é mostrada na Figura \ref{fig:log}. Para esta implementação a quantidade de elementos lógicos é mostrada na Tabela \ref{tab1}. Os dados da tabela foram obtidos a partir da ferramenta de estatísticas do circuito no \textit{Logisim 2.7.1}.
\begin{table}[!ht]
\centering
\caption{Lista de componentes utilizados na representação digital do mapa tenda com 8 bits. Utiliza-se o conceito de elementos lógicos, envolvendo portas lógicas, multiplexadores e flip-flops.}
\vspace{0.2cm}
\begin{tabular}{lc}
\hline
Componente&  Quantidade\\ \hline
     Porta XOR & 8 \\
     Multiplexador (8 bits) & 1\\
     Flip-flop D & 8\\\hline
     \textbf{Total}& 17  \\ \hline
\end{tabular}
\label{tab1}
\end{table}

A quantidade de portas \textit{XOR} e \textit{Flip-flops D} é igual a quantidade de bits do sistema. Logo um sistema de n bits terá $2n + 1$ componentes, onde $2n$ representa a quantidade de portas \textit{XOR} e \textit{Flip-flops D} e a soma de 1 representa o multiplexador com entrada de $n$ bits. Comparado com trabalhos que descrevem circuitos digitais caóticos, como \citep{Silva2017}, o circuito apresentado é significativamente mais simples, afirmando a eficiência da representação numérica por ponto fixo polarizado.

\begin{figure}[h]
    \centering
    \includegraphics[width=0.9\linewidth]{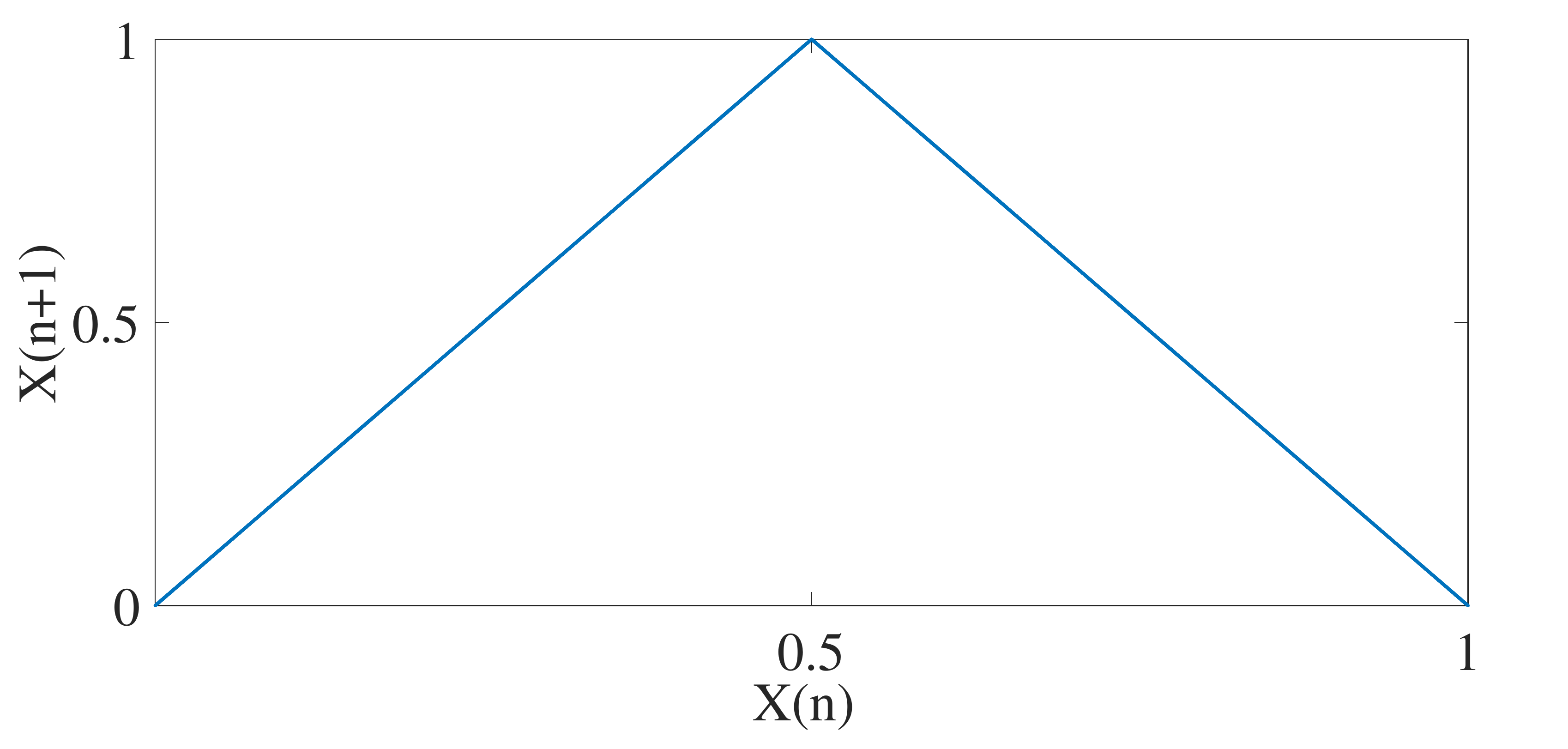}
    \caption{O mapa de primeiro retorno relaciona o valor atual $X(n)$ com o valor posterior $X(n+1)$ da série. A forma triangular é característica do mapa tenda. }
    \label{fig:mapretorno}
    \end{figure}

Os dados gerados foram convertidos para seu valor decimal utilizando o \textit{Matlab}. Através do cálculo do Expoente de Lyapunov, mostrado na Tabela \ref{tablyap}, e do mapa de primeiro retorno na Figura \ref{fig:mapretorno}, comprovou-se que o sistema proposto é capaz de representar o mapa tenda, mantendo seu comportamento caótico.

 \begin{table}[!ht]
\centering
\caption{Expoente de Lyapunov $\lambda$ calculado para séries de diferentes precisões de bits seguindo a metodologia proposta por \cite{Kantz1994} e \cite{RCD1993}. O valor da literatura é 0.693, mostrando que os resultados obtidos estão com boa aproximação do valor esperado. Como todos os valores são positivos, a representação digital do mapa tenda pode ser considerada caótica.}
\vspace{0.2cm}
\begin{tabular}{lc}
\hline
bits &  $\lambda$\\ \hline
     8 & 0,6541 \\
     16& 0,6833\\
     32 &0,6636\\\hline
\end{tabular}
\label{tablyap}
\end{table}

\begin{table*}[!ht]
\centering
\caption{Comparação da quantidade de elementos digitais de um \textit{PRNG} em diferentes trabalhos. Na quarta coluna apresenta-se a razão entre o número de elementos utilizados e a quantidade de bits. Fica evidente que o número de componentes utilizados neste trabalho é significativamente inferior a outros trabalhos da literatura. }
\vspace{0.2cm}
\begin{tabular}{l c c c}
\hline
\textit{Trabalho}&bits & Nº de Elementos & Razão\\\hline
Este trabalho &16&33 & 2,063 \\
Khani e Ahmadi(2013) &10&55 &5,500 \\
\multicolumn{1}{l}{\multirow{2}{*}{Este trabalho}} &32&65 & 2,031\\
&64&129 & 2.016\\
Sreenath e Narayanan (2018)&32&161 & 5,031\\
Sreenath e Narayanan (2018)&64&321&5,016\\ \hline
\end{tabular}
\label{tabcomp}
\end{table*}
\vskip 1mm

\begin{figure}[!ht]
    \centering
    \includegraphics[width=\linewidth]{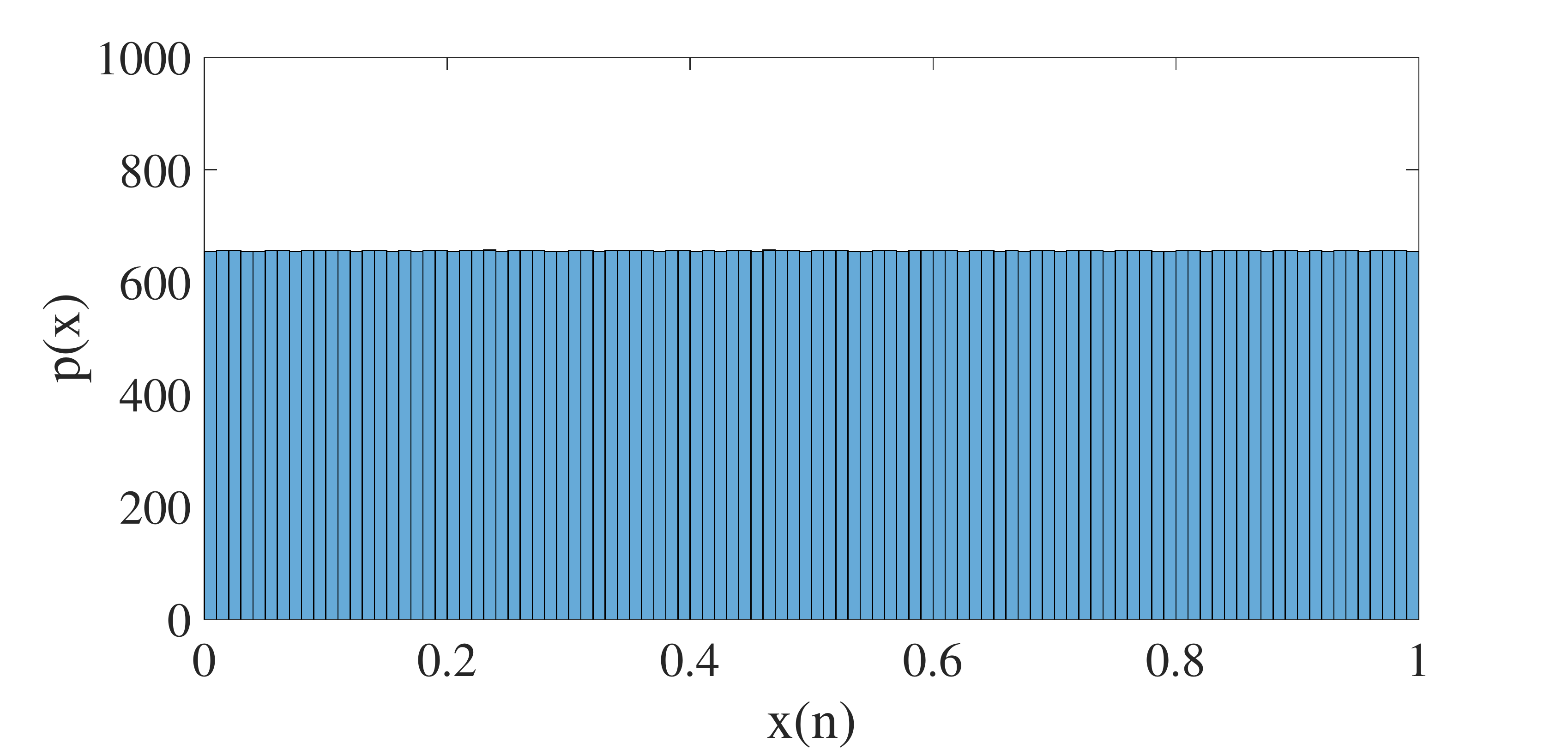}
    \caption{Histograma simulado no \textit{Matlab} de uma série de 16 bits com $2^{16}$ elementos. Percebe-se claramente a uniformidade deste histograma, indicando boas propriedades aleatórias para a representação digital do sistema.}
    \label{fig:hist}
    \end{figure}

\begin{figure}[!ht]
    \centering
    \includegraphics[width=1\linewidth]{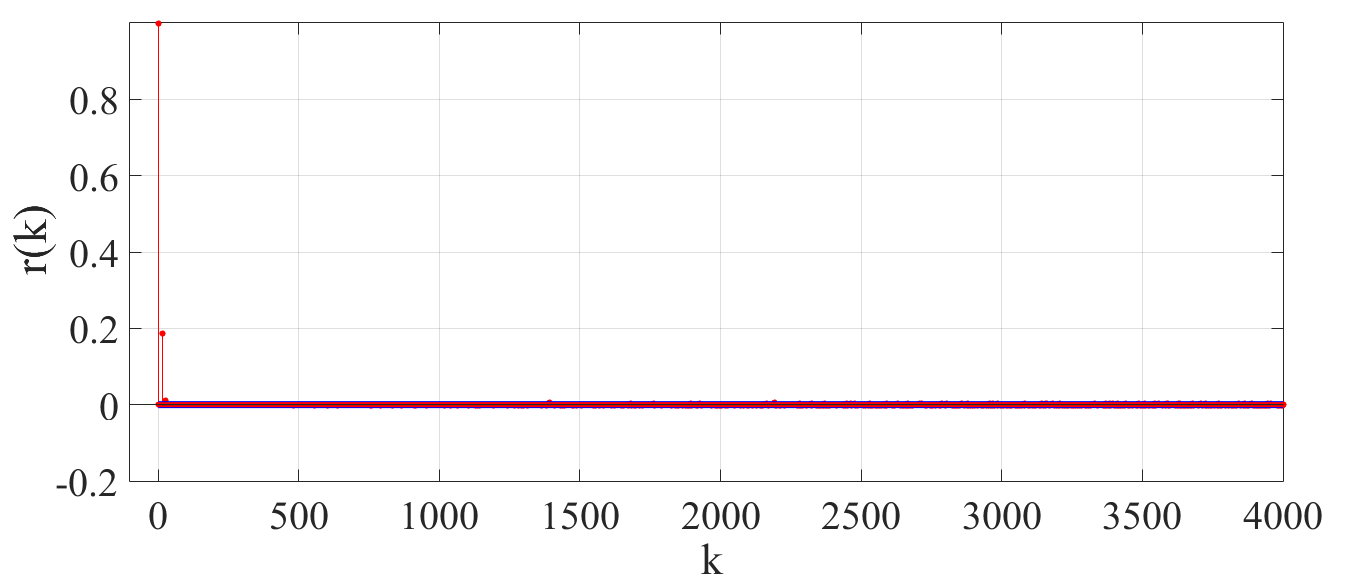}
    \caption{Autocorrelação de uma série de 16 bits com $2^{16}$ elementos, onde $k$ é o atraso e $r(k)$ é a correlação com a série atrasada. A correlação elevada somente com $k=0$ indica aperiodicidade, fator importante para um \textit{PRNG}.}
    \label{fig:autocorr}
    \end{figure}

    O desempenho do circuito como \textit{PRNG} foi avaliado através do histograma na Figura \ref{fig:hist} e da autocorrelação na Figura \ref{fig:autocorr}. O histograma mostra uma distribuição bem uniforme da série gerada e a análise da saída binária mostra um grau de desbalanceamento muito baixo, ou seja, a quantidade de 0 é muito próxima da quantidade de 1. Isso indica a elevada entropia, que tem o valor igual a $H(x) = 0,9998$.

A Tabela \ref{tabcomp} faz uma comparação da quantidade de elementos digitais  do circuito proposto com outros \textit{PRNGs}, baseados no mapa tenda, encontrados na literatura. Todos os trabalhos apresentados passaram no teste de autocorrelação. A quarta coluna mostra a razão entre o número de componentes e a quantidade de bits. Essa razão mostra que este trabalho apresenta maior eficiência comparado aos outros, conseguindo reduzir com sucesso a quantidade de componentes em mais de 50\%.

\section{Conclusão}
\label{sec:con}

Este trabalho investigou a representação digital do mapa tenda. Procurou-se nesta representação reduzir o número de componentes e evitar a degradação das propriedades caóticas deste sistema. A representação por ponto fixo polarizado permitiu uma grande redução na quantidade de elementos digitais do mapa tenda, a partir da simplificação das operações aritméticas. O sistema apresentou comportamento caótico, sendo viável para geração de números pseudo aleatórios e, portanto, para diversas aplicações. Logo a síntese de um sistema caótico \textit{PRNG} mais simplificado e eficiente foi alcançada. 

O próximo passo será a implementação desse projeto em FPGA, plataforma flexível para montagem de circuitos digitais, e a aplicação do circuito projetado na criptografia e controle, tal como feito por \cite{NEPOMUCENO201962}.



\bibliography{19sbai-mariana1}
\end{document}